# Differentiation of Distinct Single Atoms via Multi-Defocus Fusion Method

Yangfan Li [a,b], Yue Pan [a,b], Xincheng Lei [a,b], Weiwei Chen [a,b], Yang Shen [a,b], Mengshu Ge [a], Xiaozhi Liu [a], Dong Su [a,b],*

[a]Beijing National Laboratory for Condensed Matter Physics, Institute of Physics, Chinese Academy of Sciences, Beijing, 100190, China
[b] School of Physical Sciences, University of Chinese Academy of Sciences, Beijing, 100049, China

**Abstract**

High-angle annular dark-field scanning transmission electron microscopy (HAADF-STEM) is a vital tool for characterizing single-atom catalysts (SACs). However, reliable elemental identification of different atoms remains challenging because the signal intensity of HAADF depends strongly on defocus and other imaging parameters, potentially ruining the Z-contrast of atoms at different depths. In this work, we investigated the influence of the vertical position of atoms (defocus), support thickness, interatomic height, convergence and collection angles via multi-slice simulations on a model system of Fe/Pt atoms on amorphous carbon supports. Our calculation shows that at a convergence angle of 28 mrad, a defocus of 4.6 nm can cause Fe and Pt atoms indistinguishable. At a larger convergence angle, this critical indistinguishable defocus can be even shorter. To address this limitation, we propose a Multi-Defocus Fusion (MDF) method, retrieving the Z-contrast from serial images from multiple defocus. Experimental validation on a Fe/Pt SAC sample confirms the effectiveness of MDF, yielding clearly separated intensity histograms corresponding to Fe and Pt atoms. This work presents a robust, easy-to-implement strategy for accurate single-atom identification, offering valuable guidance for the accelerated screening and rational design of high-performance SACs.

*Keywords:* HAADF-STEM, Single-Atom Catalysts, Amorphous Carbon, Depth of Field, Multislice Simulation

## 1. Introduction

Single-atom catalysts (SACs) have attracted increasing attention in recent years due to their remarkable catalytic performance and maximal atomic efficiency [1-5]. Unlike conventional nanoparticle or bulk catalysts, SACs exhibit unique atomic configurations that are often decisive in determining catalytic activity and selectivity across a range of chemical systems [6-9]. Comprehensive characterization at the atomic scale—identifying the elemental species, spatial distribution, and coordination environment—is therefore essential for understanding and optimizing SAC performance [9-12]. Advancements in scanning transmission electron microscopy (STEM) have enabled the direct imaging of individual atoms on various substrate materials [13, 14]. In particular, high-angle annular dark-field STEM (HAADF-STEM) has emerged as a critical tool in SAC characterization, offering both high spatial resolution and atomic number sensitivity (Z-contrast) [15]. When combined with techniques such as extended X-ray absorption fine structure (EXAFS), HAADF-STEM has become a gold standard for verifying the synthesis of SACs, including single-atom, dual-atom, and multi-atomic configurations. For example, recent studies have exploited HAADF-STEM to directly visualize Fe and Pt single atoms on carbon supports, with image simulations clarifying how imaging conditions and sample structure affect atomic detectability [16, 17]. Liu *et al.* [18] employed an atom-by-atom counting approach to precisely identify the active sites in the Pt single-atom catalyst. Lin *et al.* [19] conducted a comprehensive investigation that not only elucidated the generation and dispersion mechanisms of Fe single atoms on their supports but also systematically analyzed their distribution and chemical states, thereby establishing critical connections between atomic configurations and catalytic performance.

Despite its widespread use, HAADF-STEM imaging of SACs remains sensitive to both experimental conditions and sample characteristics. While atoms with significantly different Z values are generally distinguishable, this Z-contrast can be compromised by variations in the vertical position of atoms (defocus), support thickness, and interatomic height along the beam direction. As shown in **Fig. 1**, our simulations demonstrate that these parameters significantly affect the signal intensity from single atoms. In this case, the background-subtracted signal (BSS) does not scale with a power function of atomic number Z, complicating direct elemental identification. This challenge is exacerbated in cases where SACs contain different atomic species distributed at varying depths, making reliable distinction between elements like Pt and Fe difficult or even impossible. Although complementary techniques such as STEM electron energy-loss spectroscopy (STEM-EELS) can provide elemental information, they suffer from prolonged acquisition time, beam damage artifacts, and increased sensitivity to specimen drift and single-atom displacements—rendering them unsuitable for high-throughput SAC analysis [20]. At present, how to reliably distinguish different single-atom species and disentangle the intertwined effects from imaging conditions and sample geometry remains a formidable challenge.

---

\* Corresponding author.
 *E-mail address:* dongsu@iphy.ac.cn.

In this study, we address this challenge by combining rigorous image simulations with experimental validation. We systematically investigate the influence of imaging parameters—including defocus, convergence and collection angles, and sample configuration—on the BSS of individual atoms in SACs. Using a model system of Pt and Fe single atoms dispersed on an amorphous carbon support, we demonstrate that under certain conditions (e.g., 4.6 nm vertical interval distance and 28 mrad convergence semi-angle), Pt atoms can become indistinguishable from Fe atoms. To overcome these limitations, we propose a simple yet effective Multi-Defocus Fusion (MDF) method, which integrates multiple HAADF-STEM images acquired at different defocus values. Experimental results confirm that the MDF method successfully resolves the distribution of Pt and Fe atoms using just three images. This simple method offers a robust approach for enhancing single-atom discrimination in multi-element SACs.

**2. Materials and methods**

*2.1 Multi-slice Theory and abTEM Simulations*

HAADF-STEM imaging provides atomic-resolution elemental contrast by detecting electrons elastically scattered at high angles. According to Rutherford scattering theory, the differential scattering cross-section is given by[21]:

$$\frac{d\sigma}{d\Omega} = \left(\frac{Ze^2}{16\pi\varepsilon_0 E}\right)^2 \frac{1}{sin^4(\theta/2)} \qquad (1)$$

This implies an ideal intensity dependence proportional to $Z^2$. However, in practice, factors such as electron cloud screening, detector geometry, thermal vibrations (Debye-Waller factors), and multiple scattering events reduce the exponent to a typical experimental range of 1.2–1.8 [22]. Despite these deviations, HAADF-STEM remains a powerful method for imaging heavy atoms dispersed on light support.

The multi-slice method is a well-established computational approach for simulating high-resolution TEM and STEM images [23-25]. It divides the specimen into a series of thin slices along the electron beam direction. Mathematically, the propagation of the wavefunction $\psi$ from slice *n* to *n+1* can be expressed as [26]:

$$\psi_{n+1}(x,y) = F^{-1}\{F[\psi_n(x,y) \cdot t_n(x,y)] \cdot p(k_x, k_y)\} \qquad (2)$$

Here, $t_n(x,y)$ represents the slice transmission function, and $p(k_x, k_y)$ is the propagator function in reciprocal space. This formulation, efficiently implemented via Fast Fourier Transform (FFT), provides accurate modeling of electron-specimen interactions at the atomic scale.

All simulations in this study were performed using **abTEM**, a Python-based open-source platform designed for electron microscopy simulations based on multi-slice theory [27]. All image simulations were performed assuming an electron microscope operating at 200 kV acceleration voltage. The simulations systematically varied the convergence semi-angle (20 mrad, 28 mrad, and 36 mrad) and the HAADF detector collection angle (45–250 mrad, 54–250 mrad, and 68–250 mrad) (see **Table S1** for details). In our simulation, we choose a changing the spherical aberration coefficients ($C_s$) of 500nm based on the experimental value. We further simulate the HAADF images of Fe atom by changing $C_s$. As shown in **Fig. S1**, the increase of $C_s$ causes the increase of probe size and thus the full width at half maximum (FWHM) of HAADF signal.

*2.2 Structural Models of SACs*

The simulated SAC structures consist of isolated Fe and Pt atoms dispersed on an amorphous carbon support [28, 29]. Amorphous carbon models were generated using a "C-cube" method, starting from crystalline carbon slabs whose atomic positions were randomly displaced by ~30% of the typical interatomic spacing to mimic amorphous disorder. These structures were validated against experimental EELS-based thickness measurements[30], providing a realistic basis for image simulation. **Fig. S2** provides EELS measurements of the actual support thickness used in corresponding experimental imaging.

To evaluate the effects of substrate thickness and defocus on image contrast, we constructed models with single Fe or Pt atoms placed centrally on amorphous carbon supports of varying thickness (0–10 nm). This setup allowed for detailed analysis of single-atom visibility, contrast variation, and depth resolution across different imaging conditions.

To simulate more realistic atomic distributions, two additional configurations were considered:
- **Surface Distribution Model**: 100 Fe and 100 Pt atoms were randomly distributed within the uppermost and lowermost 1 nm surface layers of a 7 nm-thick amorphous carbon slab, mimicking surface-bound atom arrangements.
- **Uniform Distribution Model**: 100 Fe and 100 Pt atoms were randomly embedded throughout the entire volume of the same 7 nm-thick support, representing SACs with homogeneously dispersed atoms.

These complementary models enabled evaluation of our depth-sectioning strategy under diverse atomic configurations.

*2.3 Experimental details*

HAADF-STEM experiments were carried out using a JEOL NEOARM transmission electron microscope operating at 200 kV. In consistent with simulation results, the convergence semi-angle was set to 28 mrad, and the HAADF detector collection angle spanned 54–250 mrad to enhance atomic-number contrast and optimize single-atom visibility.

Samples were prepared by drop-casting a carefully diluted solution containing $FeCl_3$ and $H_2PtCl_6$ in a 1:1 molar ratio onto ultrathin amorphous carbon-coated TEM grids. The solution concentration was optimized to yield isolated atomic dispersion while minimizing possible agglomeration. After deposition, the grids were allowed to dry under ambient conditions to preserve atom separation.

*2.4 Multi-Defocus Fusion method*

The Multi-Defocus Fusion (MDF) method enables enhanced elemental discrimination by accurately identifying and correlating single atoms across HAADF-STEM images acquired at different defocus values. This approach requires a robust computational workflow to extract the maximum atomic intensities for subsequent statistical analysis.

To achieve this, we first detect atomic positions in each image using the Laplacian of Gaussian (LoG) algorithm[31], a proven technique widely employed in electron microscopy for the identification of atomic-scale features. The LoG operator combines Gaussian smoothing with a second derivative operation, effectively highlighting local intensity maxima indicative of atomic positions[32]. Mathematically, the LoG operation can be expressed as:

$$\nabla^2 G(x, y; \sigma) = \frac{x^2 + y^2 - 2\sigma^2}{\pi \sigma^6} exp\left(-\frac{x^2 + y^2}{2\sigma^2}\right) \tag{3}$$

where $\nabla^2 G(x, y; \sigma)$ denotes the Gaussian function with standard deviation $\sigma$, and $\nabla^2$ is the Laplacian operator. Appropriate selection of the Gaussian kernel width ($\sigma$) allows optimized detection sensitivity and accuracy for single-atom positions.

Next, to correlate atomic positions across the defocus series, we employ the Lucas-Kanade optical flow algorithm[33]. This method assumes both brightness constancy and local spatial coherence between sequential images. For an atomic position point $p = (x, y)$ in the first image, its displacement vector $d = (u, v)$ in the subsequent image frame is estimated by minimizing the following error function within a local neighbourhood $W$:

$$E(u,v) = \sum_{(x,y) \in W} [I_1(x,y) - I_2(x+u, y+v)]^2 \qquad (4)$$

where $I_1$ and $I_2$ represent the intensity distributions of two consecutive defocus images.

By combining accurate atom detection using LoG with reliable tracking via optical flow, this computational strategy ensures robust identification and correspondence of single atoms across all focal planes, thereby constituting the core of the MDF algorithm.

## 3. Results and Discussion

*3.1 Simulations on the Effect of Defocus and Support Thickness*

The defocus setting and the thickness of the amorphous carbon support play crucial roles in determining the visibility of individual atoms in STEM-HAADF imaging. We then performed simulations by varying these two parameters as presented in **Fig. 1**. **Fig. 1(a)** is a schematic of the model used for the simulation, with Fe atoms on amorphous carbon support. **Fig. 1(b)** presents the simulated HAADF-STEM images with varying thicknesses (0~10 nm) and defocus values (-5 nm~ +5 nm). Under conditions where the electron beam is precisely focused at the atomic plane of the Fe atom (defocus = 0 nm), the Fe atom exhibits the highest image intensity, maximizing the atom's visibility. To quantitatively evaluate their signal intensity, line intensity profiles across the position of the Fe atom were extracted. **Fig. 1(c)** shows profiles at a fixed carbon support thickness of 7 nm under varying defocus positions, and **Fig. 1(d)** illustrates the intensity profiles at a constant defocus of +1 nm but with varying carbon thicknesses. Our simulations demonstrate the significant impact of defocus on the detectability of single Fe atoms. As the focal plane shifts away from the atom's position, either above or below it, the peak intensity decreases significantly. Particularly, when the focal plane shifts into the amorphous carbon support, fluctuations in local carbon density can form bright contrast maxima, closely resembling single atom signals and thus complicating accurate atom identification. Consequently, maintaining appropriate defocus conditions is critical to reliably distinguish single Fe atoms from background noise in the amorphous carbon support.

Furthermore, as revealed in **Fig. 1(d)**, the thickness of carbon support also influences the visibility of Fe atoms. Initially, with the increase of carbon thickness, the signal of Fe atoms increases due to additional scattering from carbon support. At the same time, the scattering intensity of carbon support also increases. The increasing background noise from the amorphous carbon support leads to a reduction in the BSS of Fe atoms. However, compared to the defocus, the thickness of carbon support plays a secondary role on the HAADF signal of Fe.

To further investigate the influence of different support materials on HAADF imaging of different kinds of atoms, we performed analogous simulations on Pt atom on either carbon or silicon support as well as Fe atom on silicon support, as shown in **Fig. S3-S5**, respectively. Under the same conditions, the simulations show distinguishing single Fe atoms on amorphous silicon proves to be much more challenging while imaging Pt is much easier, in consistent with the anticipation.

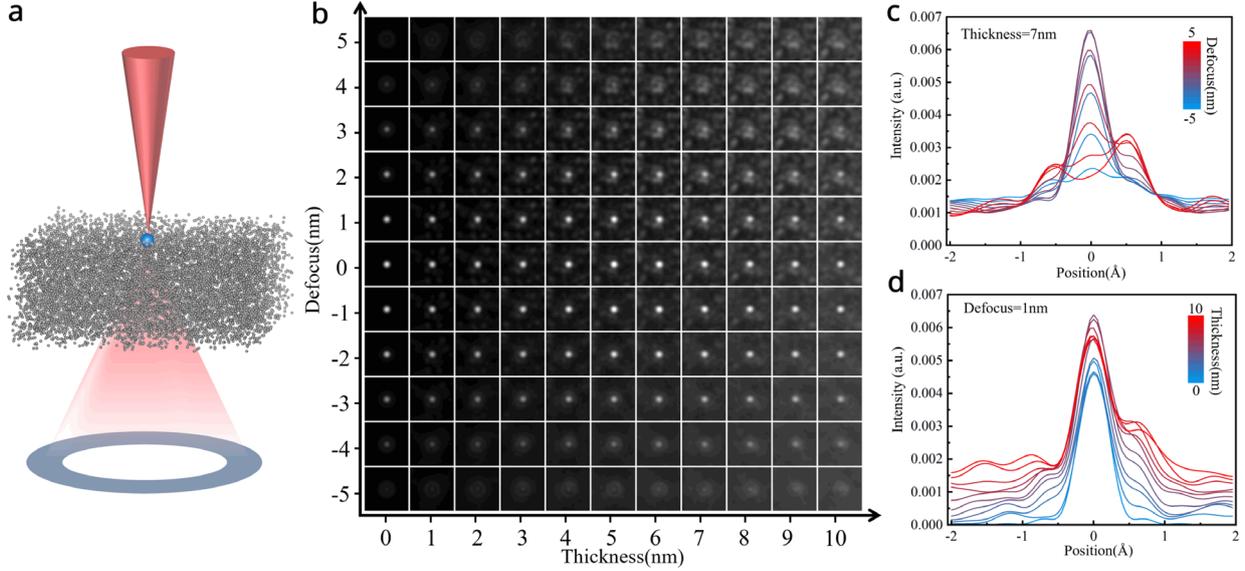

**Fig. 1**. HAADF-STEM imaging simulation of single Fe atoms on amorphous carbon supports. (a) Schematic illustration of the HAADF-STEM mode and the Fe-Carbon model. (b) Simulated map images obtained under various amorphous carbon thicknesses (0–10 nm) and defocus (−5 to +5 nm) conditions. (c) Line scans recorded at a fixed carbon thickness of 7 nm under different defocus values. (d) Line scans recorded at a fixed defocus of 1 nm for amorphous carbon supports of varying thickness.

*3.2 Simulations on the Effect of Convergence and Collection Angles*

In SAC systems containing different atomic species, such as Pt and Fe, the ability to distinguish between atomic numbers via Z-contrast is essential. In the case of single Fe and Pt atoms on amorphous carbon support, a typical kind of dual-atom SAC [34, 35], we investigated the impact of convergence and collection semi-angles on the Z-contrast of HAADF-STEM image, as shown in **Fig. 2 and Fig. S6**. **Fig. 2(a)** presents simulated HAADF-STEM contrast values for single Fe atoms dispersed on amorphous carbon supports of varying thicknesses under a series of varied convergence and collection angles. For these simulations, the image contrast was quantified as [36, 37]:

$$Contrast = \frac{I_{max} - I_{min}}{I_{min}} \tag{5}$$

where $I_{max}$ denotes the maximum intensity at the atomic center and $I_{min}$ indicates the average intensity of the surrounding background region. This definition provides a normalized indication of atomic visibility against the background, facilitating quantitative comparisons between different imaging conditions. **Fig. 2(a)** demonstrates that the simulated HAADF-STEM contrast for single Fe atoms is notably enhanced with thinner amorphous carbon supports and larger convergence semi-angles, which sharpen the distinction between atom and background.

The influences of convergence and collection angles on the resulting HAADF contrast for single Fe atoms were further analyzed in **Fig. 2(b)** and **2(c)**, respectively. **Fig. 2(b)** indicates that at a fixed collection angle range of 68–250 mrad, the contrast enhances with the increased convergence angles. This phenomenon can be attributed primarily to the reduction in the depth of field (DOF) associated with larger convergence angles, as larger angles increase the beam's axial resolution. Our simulations indirectly corroborate an approximate formula for DOF [38, 39]. The transverse probe size ($d_{xy}$) and the axial resolution ($d_z$) are approximately related to the convergence angle ($\alpha$), electron wavelength ($\lambda$), spherical aberration ($C_s$), chromatic aberration ($C_c$), brightness ($B$), probe current ($I_P$), and energy spread ($\Delta E/E$) by the following equations [38, 39]:

$$d_{xy} = \sqrt{(2\pi\alpha\sqrt{\frac{I_p}{B}})^2 + (\frac{1}{2}C_s\alpha^3)^2 + (\frac{0.6\lambda}{\alpha})^2 + (C_c\alpha\frac{\Delta E}{E})^2} \tag{6}$$

$$d_z \approx \frac{d_{xy}}{\alpha} \approx \frac{\lambda}{\alpha^2} \tag{7}$$

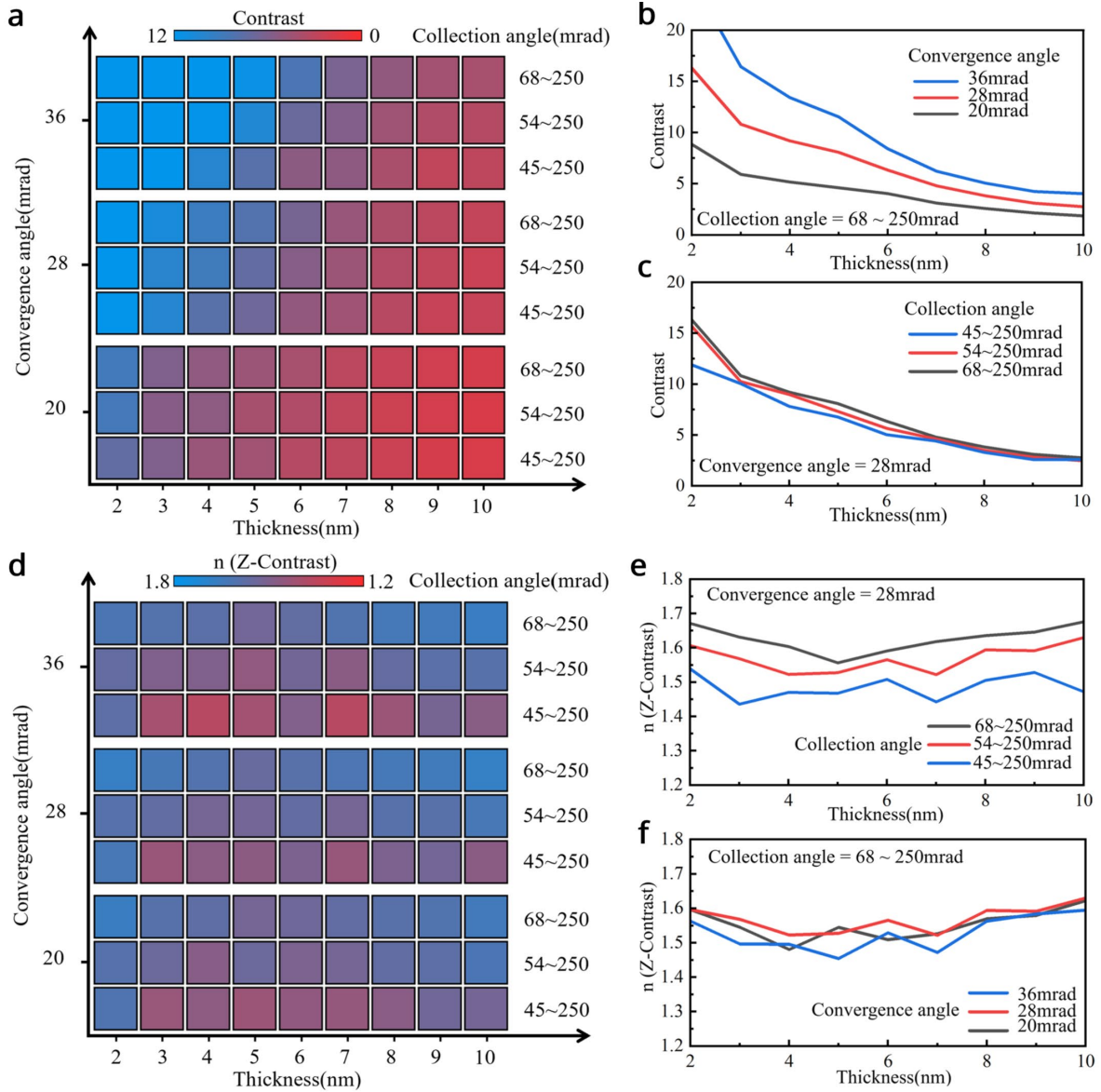

**Fig. 2**. Simulation on the HAADF-STEM contrast and elemental discrimination capability of Pt and Fe atoms on amorphous carbon support under varying convergence and collection angles. (a) Contrast maps for Fe atoms on amorphous carbon supports of different thicknesses acquired with various convergence and collection angles. (b) Dependence of Fe contrast on different convergence angles at a collection angle of 68–250 mrad. (c) Dependence of Fe contrast on different collection angles at a convergence angle of 28 mrad. (d) Color maps of *n* values, as calculated from the Z-contrast scaling relationship of Eq.8 under different convergence and collection angles, respectively. (e) Dependence of *n* on different collection angles at a fixed convergence angle of 28 mrad. (f) Dependence of *n* on varying convergence angles at a collection angle of 68–250mrad. The defocus is set at 0 nm for all simulations.

Here, smaller DOF confines signal collection to a narrow focal plane, improving atomic contrast. Consequently, the contrast of single Fe atoms against the amorphous carbon background is notably enhanced under conditions of increased convergence angles.

**Fig. 2(c)** shows that collection angle variation has a subtle effect on contrast. Although larger collection angles theoretically increase Z-contrast, smaller angles can improve signal-to-noise ratio by collecting more scattered electrons. Therefore, practical imaging requires balancing contrast enhancement with noise reduction.

To evaluate elemental distinguishability, we computed the Z-contrast scaling exponent $n$ between Pt (Z=78) and Fe (Z=26) using:

$$n = log_{78/26} \frac{I_2}{I_1} \tag{8}$$

where $I_2$ and $I_1$ represent the HAADF signal intensities for single Pt and Fe atoms, respectively. This exponent $n$ effectively quantifies the elemental sensitivity.

**Fig. 2(d)** shows the calculated $n$ factor for different convergence and collection angles across amorphous carbon support with varying thicknesses. Our calculation demonstrates that $n$, the elemental discrimination capability, varies considerably with the imaging conditions. **Fig. 2(e)** and **2(f)** present how convergence and collection angles affect the exponent $n$, respectively. **Fig. 2(e)** highlights a strong dependence of $n$ on the inner and outer collection angles at a fixed convergence angle of 28 mrad. Larger collection angles consistently lead to a higher $n$-value, corresponding to an enhanced elemental distinction between Pt and Fe. In contrast, **Fig. 2(f)** specifically explores the impact of changing convergence angles at a fixed collection angle range of 68–250 mrad. It is shown that variations in the convergence angle have a minor influence on the value of $n$, suggesting that elemental discrimination in HAADF-STEM imaging is relatively insensitive to the convergence angle. This observation agrees with the established theoretical predictions for Z-contrast imaging, which emphasize improved elemental sensitivity at higher scattering angles due to the increased significance of Rutherford scattering.–

*3.3 Differentiation between Pt and Fe Atoms at Different Heights*

In SACs with Fe and Pt single atoms on carbon support, individual Fe and Pt atoms may locate at different vertical positions relative to the focal plane. As demonstrated in **Section 3.1**, defocus significantly affects atom intensity; thus, vertical displacement can obscure elemental identity. We simulated HAADF signals from Pt and Fe atoms placed at varying heights, as shown in **Fig. 3(a)**. **Fig. 3(b–j)** present intensity curves for both elements across different convergence and collection angles. The resulting data consistently demonstrate that the primary factor influencing the vertical range over which Fe and Pt atom intensities overlap is the value of the convergence angle used during HAADF-STEM imaging. As indicated in the dashed lines in **Fig. 3(b–j)**, the brightness of Fe and Pt atoms becomes equal when the electron probe focuses on Fe atoms and Pt atoms are located at the critical defocus height named as $d_{fp}$. Beyond this Pt-Fe interval distance, the signal intensity of Pt acquired at defocus state is less than the maximum of Fe atoms, making Pt become indistinguishable from Fe in HAADF contrast. As convergence angle increases, $d_{fp}$ decreases due to the shrinking DOF. Narrower DOF leads to a rapid decline in atom intensity even for small deviations from the exact focal plane.

To generalize this phenomenon across elements, we simulate the HAADF signals for single atoms with atomic numbers (Z) spanning from Na (Z=11) to Pb (Z=82) as shown in **Fig. 3 (h)**. We can observe, across a broader range of elements, that the signal of HAADF is highly affected by the height. With a defocus of 8 nm, it is even hard to distinguish the light element of Na from the heavy element of Pb. With the dashed line denoting the in-focus brightness of Fe atoms, the indistinguishable defocus distance $d_{fp}$, can be plotted with a function of atomic number, as shown in **Fig. 3(i)**. The curve can be well-fitted using a logarithmic function. The best-fit equation is $d_{fp}$ = 1.174 $In$ (Z – 24.699), with an R² value of 0.99. It is common that SACs contain different kinds of atoms at different height. These results highlight a critical challenge: even heavy elements can be indistinguishable from lighter ones if imaged out of focus.

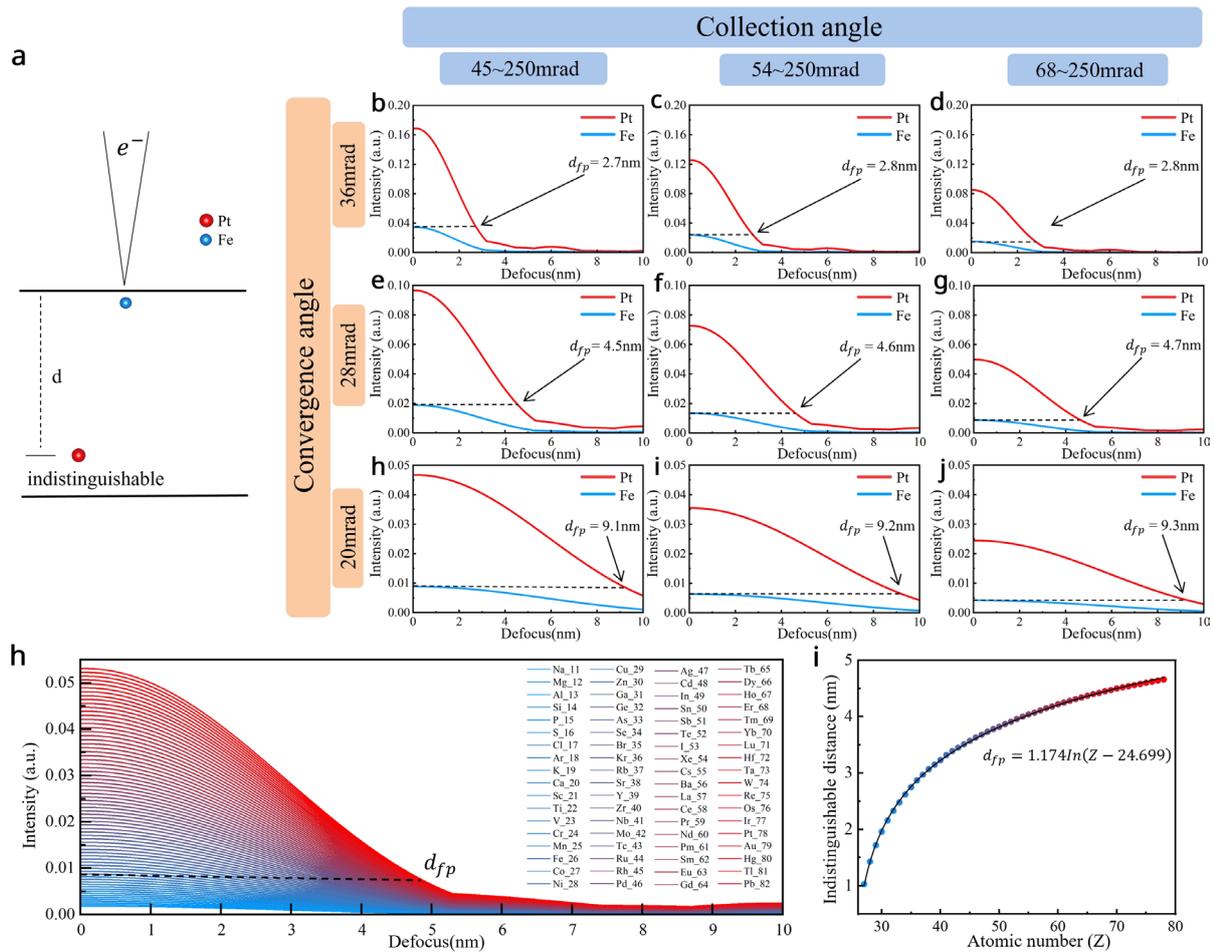

**Fig. 3.** Simulation on the signals from single Fe and Pt atoms under varying imaging conditions. (a) Schematic illustration of the simulation model. (b)–(j) Plots of the brightness of single Pt and Fe atoms versus defocus at different convergence and collection angles. The intersections of the dashed curves indicate the height difference as the equal brightness. (h) Simulated HAADF signals for single atoms at different defocus with atomic numbers ranging from Z = 11 to Z = 82 under the imaging conditions used in (g); the dashed line indicates the in-focus brightness of Fe atoms. (i) Plot of $d_{fp}$ as a function of atomic number, obtained by extracting the horizontal coordinates at which the dashed line in (h). The fitting curve has a value of $R^2$=0.99.

To overcome the ambiguity introduced by vertical atomic displacement, we propose here a Multi-Defocus Fusion (MDF) method. As discussed in section 2.4, this methodology retrieves the signal intensity of each atom from selecting the maximum value from multiple images acquired at different defocus conditions, leveraging a close imitation to the ideal in-focus Z-contrast. We simulate two types of SACs: **Fig. 4(a)** depicts Fe and Pt atoms uniformly dispersed within a 7 nm-thick amorphous carbon support, while **Fig. 4(i)** represents a scenario with Fe and Pt atoms randomly positioned on the upper and lower surfaces of the carbon support. Each model contains 100 Fe atoms and 100 Pt atoms. We simulated HAADF-STEM images for these two types of models, using a convergence angle of 28 mrad and a collection angle of 54–250 mrad. All images were recorded with an image resolution of 1024 × 1024 pixels, corresponding to a pixel size of 0.033 nm, thus providing sufficient spatial resolution to reliably resolve individual atoms. To balance image quality and minimize possible beam-induced damage or atom mobility during acquisition, the pixel dwell time was fixed at 6 μs.

Simulated HAADF-STEM images are presented with three different defocus — at the upper surface (**Fig. 4(b)** and **(j)**), at the middle plane (**Fig. 4(c) and (k)**), and at the lower surface (**Fig. 4(d) and (l)**), respectively. To visualize the signal of Fe and Pt atoms, we applied contrast inversion to the simulated images, and the original images are provided in the Supporting Information (**Fig. S7**). As expected, atomic signal intensities shift with defocus. The same atom appears brighter in one focal plane and dimmer in others, depending on its vertical position.

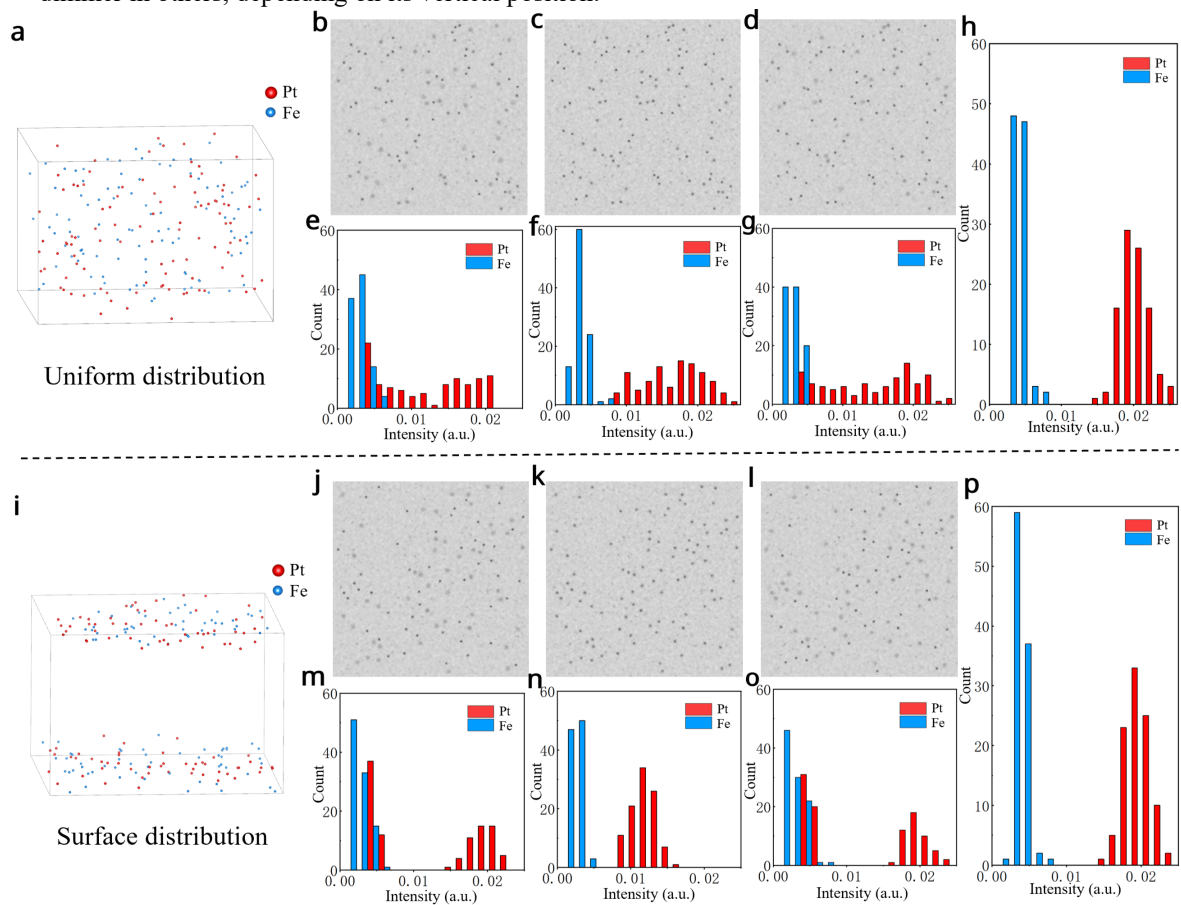

**Fig. 4**. Different statistical distribution of Fe and Pt atoms on a 7 nm-thick amorphous carbon support. (a) Uniform distribution model of 100 Fe and 100 Pt atoms within the interior of a 7 nm-thick amorphous carbon support. (b–d) Simulated images (inverse contrast) focused on the top, middle, and bottom regions, respectively. (e–g) Intensity distribution maps for Pt and Fe corresponding to the simulated images. (h) Composite histogram obtained by superimposing the highest atomic intensities from the three simulated regions. (i) Surface distribution model depicting 100 Fe atoms and 100 Pt atoms on the support. (j–l) Simulated images (inverse contrast) focused on the top, middle, and bottom regions, respectively. (n–o) Corresponding intensity distribution maps for Pt and Fe at the respective locations. (p) Composite histogram generated by overlaying the highest atomic intensities from the three simulated regions.

The histograms of these simulated HAADF-STEM images are plotted below the images. Statistical intensity distributions for Fe and Pt atoms from each defocus condition are presented in **Figs. 4(e–g)** (uniform internal distribution) and **Figs. 4(m–o)** (surface distribution). Since the data is from simulation, we can assign atoms to either Fe or Pt. Noticeable overlaps in intensity distributions between Fe and Pt atoms exist in each single-defocus image, confirming that Fe and Pt atoms cannot be distinguished from a single HAADF-STEM image. However, by applying the intensity maximization strategy—simply selecting the maximum intensity value for each identified atom across the three images—we achieve markedly improved elemental differentiation, as evidenced in **Fig. 4(h)** and **Fig. 4(p)**. MDF integrates these images to reconstruct a more complete view of atom identities, positions and heights, dispelling defocus-induced ambiguity.

*3.4 Experimental Verification*

To experimentally validate the MDF method, we performed HAADF-STEM imaging on a SAC specimen containing single Fe and Pt atoms on amorphous carbon support, using a Cs-corrected JEOL NEOARM TEM. **Fig.s 5(a–c)** illustrate representative HAADF-STEM images of the supported single atoms acquired at three distinct defocus conditions: upper focus (**Fig. 5a**), middle focus (**Fig. 5b**), and lower focus (**Fig. 5c**), respectively (The original experimental images are shown in **Fig. S8**). These chosen defocus settings allowed us to compare our experimental results with the simulations of **Fig. 4**. Intensity distribution histograms of **Fig. 5d–f** are generated individually from defocus images of **Fig. 5a–c**, respectively. These histograms demonstrate that the differentiation between Fe and Pt atoms based solely on single-image intensity distributions are unworkable.

We then implement a multi-defocus fusion (MDF) method, where atomic positions were tracked and correlated across the three defocus images with optical flow technique (**Fig. 5a–c**). For each atomic site, the maximum observed intensity across the three focal planes was recorded and used to construct a combined intensity histogram, as shown in **Fig. 5(g)**. Remarkably, this integrated statistical analysis reveals two well-separated intensity peaks, which can be unambiguously assigned to Fe and Pt atoms. The lower-intensity peak is attributed to Fe atoms, whereas the higher-intensity peak corresponds to Pt atoms, consistent with their respective atomic numbers and the underlying Z-contrast mechanism of HAADF-STEM imaging. These experimental results confirm the effectiveness of the MDF method in enhancing elemental discrimination of single atoms, even in complex imaging environments.

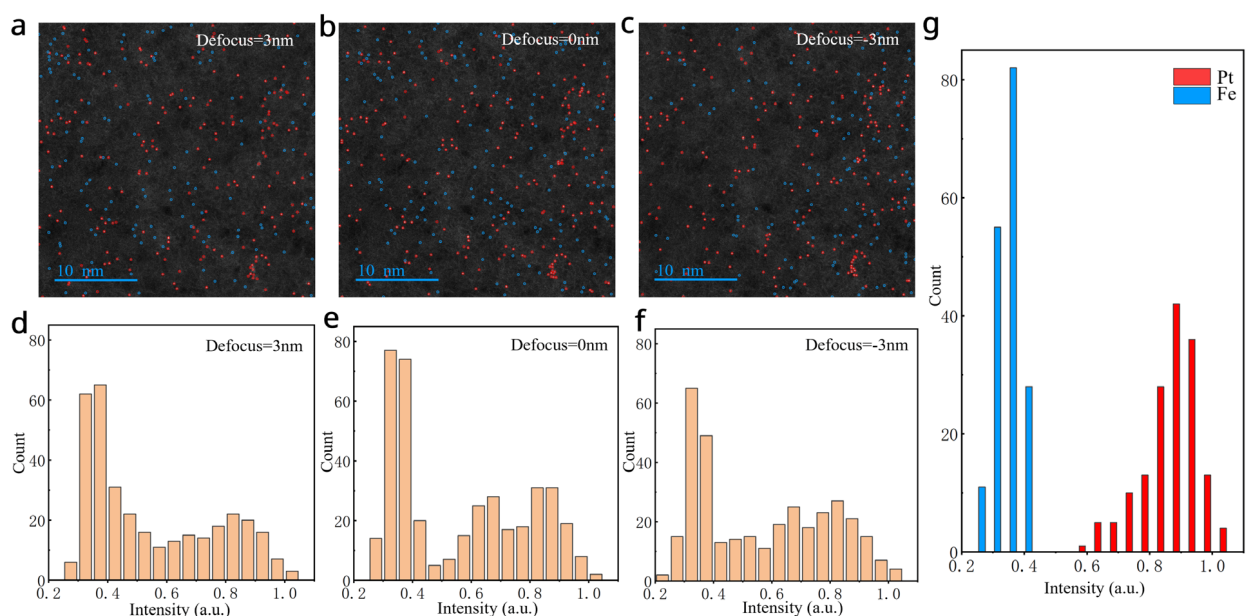

**Fig. 5**. Experimental results of STEM-HAADF images of Fe and Pt single atom catalysts on amorphous carbon support. (a–c) HAADF images acquired at different electron beam defocus settings: upper (a), middle (b), and lower (c). (d–f) Corresponding intensity distribution histograms obtained from atomic positions identified by the Laplacian of Gaussian (LoG) algorithm for images (a–c), respectively. (g) Statistical intensity distribution histogram of single Fe and Pt atoms obtained with MDF method.

**4. Conclusions**

In summary, we investigated how experimental parameters of HAADF-STEM imaging influence the precise characterization of single atoms with simulations on a model system of single Fe and Pt atoms on amorphous carbon support. Comprehensive multi-slice calculations revealed critical dependencies of single-atom visibility and elemental discrimination on the vertical positions of atoms(defocus), substrate thickness, convergence semi-angle, and the collection angle. Our study shows that the vertical position of atoms significantly impacts signal intensity, leading to potential misidentification. For instance, under a critical defocus height of 5 nm, the HAADF signal from a Pt atom may become indistinguishable from that of an Fe atom. Simulated HAADF signals for single atoms with atomic numbers ranging from $Z = 11$ to $Z = 82$ were also calculated as a function of defocus, highlighting the challenge of distinguishing atoms in systems containing mixed atomic species. We therefore introduced a MDF method that correlates atomic intensities across several defocused images. By extracting the maximum intensity at each site, the MDF method enhances elemental contrast and enables reliable discrimination between different types of atoms. Incorporating additional images within a defined depth range further enhances the effectiveness of the MDF method. Experimental validation on a dual-atom Fe–Pt system confirmed the method's ability to distinguish atomic species accurately, demonstrating its applicability to broader multi-species single-atom systems. Our findings offer both practical guidance and conceptual insight for the application of HAADF-STEM in single-atom analysis. The MDF method provides a straightforward yet powerful strategy to overcome depth-induced ambiguity, advancing the research of atomic-scale characterization in catalysis and materials science.

**Acknowledgements**

This work was supported by the National Key Research and Development Program of China (2023YFB4006201), the National Natural Science Foundation of China (No. U21A20328 and 22209202), and the Strategic Priority Research Program (B) (No. XDB33030200) of Chinese Academy of Sciences.

**References**

[1] X.F. Yang, A.Q. Wang, B.T. Qiao, J. Li, J.Y. Liu, T. Zhang, Single-Atom Catalysts: A New Frontier in Heterogeneous Catalysis, Accounts of Chemical Research, 46 (2013) 1740-1748.
[2] H.B. Zhang, G.B. Liu, L. Shi, J.H. Ye, Single-Atom Catalysts: Emerging Multifunctional Materials in Heterogeneous Catalysis, Advanced Energy Materials, 8 (2018) 1701343.
[3] A.Q. Wang, J. Li, T. Zhang, Heterogeneous single-atom catalysis, Nature Reviews Chemistry, 2 (2018) 65-81.
[4] N. Cheng, L. Zhang, K. Doyle-Davis, X. Sun, Single-Atom Catalysts: From Design to Application, Electrochemical Energy Reviews, 2 (2019) 539-573.
[5] B.T. Qiao, A.Q. Wang, X.F. Yang, L.F. Allard, Z. Jiang, Y.T. Cui, J.Y. Liu, J. Li, T. Zhang, Single-atom catalysis of CO oxidation using Pt1/FeOx, Nature Chemistry, 3 (2011) 634-641.
[6] F. Tao, P.A. Crozier, Atomic-Scale Observations of Catalyst Structures under Reaction Conditions and during Catalysis, Chemical Reviews, 116 (2016) 3487-3539.
[7] H. Yan, C.L. Su, J. He, W. Chen, Single-atom catalysts and their applications in organic chemistry, Journal of Materials Chemistry A, 6 (2018) 8793-8814.
[8] X. Cai, X. Sui, J. Xu, A. Tang, X. Liu, M. Chen, Y. Zhu, Tuning Selectivity in Catalytic Conversion of CO2 by One-Atom-Switching of Au9 and Au8Pd1 Catalysts, CCS Chemistry, 3 (2021) 408-420.
[9] S.F. Ji, Y.J. Chen, X.L. Wang, Z.D. Zhang, D.S. Wang, Y.D. Li, Chemical Synthesis of Single Atomic Site Catalysts, Chemical Reviews, 120 (2020) 11900-11955.
[10] E.D. Boyes, A.P. LaGrow, M.R. Ward, R.W. Mitchell, P.L. Gai, Single atom dynamics in chemical reactions, Accounts of Chemical Research, 53 (2020) 390-399.
[11] J. Liu, Aberration-corrected scanning transmission electron microscopy in single-atom catalysis: Probing the catalytically active centers, Chinese Journal of Catalysis, 38 (2017) 1460-1472.
[12] P. Tieu, X. Yan, M. Xu, P. Christopher, X. Pan, Directly probing the local coordination, charge state, and stability of single atom catalysts by advanced electron microscopy: a review, Small, 17 (2021) 2006482.
[13] D.A. Muller, L.F. Kourkoutis, M. Murfitt, J.H. Song, H.Y. Hwang, J. Silcox, N. Dellby, O.L. Krivanek, Atomic-Scale Chemical Imaging of Composition and Bonding by Aberration-Corrected Microscopy, Science, 319 (2008) 1073-1076.
[14] O.L. Krivanek, T.C. Lovejoy, N. Dellby, Aberration-corrected STEM for atomic-resolution

imaging and analysis, Journal of Microscopy, 259 (2015) 165-172.
[15] P.D. Nellist, S.J. Pennycook, The principles and interpretation of annular dark-field Z-contrast imaging, in: P.W. Hawkes (Ed.) Advances in Imaging and Electron Physics, Elsevier, 2000, pp. 147-203.
[16] J. Liu, M. Jiao, L. Lu, H.M. Barkholtz, Y. Li, Y. Wang, L. Jiang, Z. Wu, D. Liu, L. Zhuang, C. Ma, J. Zeng, B. Zhang, D. Su, P. Song, W. Xing, W. Xu, Y. Wang, Z. Jiang, G. Sun, High performance platinum single atom electrocatalyst for oxygen reduction reaction, Nature Communications, 8 (2017) 15938.
[17] W.S. Song, M. Wang, X. Zhan, Y.J. Wang, D.X. Cao, X.M. Song, Z.A. Nan, L. Zhang, F.R. Fan, Modulating the electronic structure of atomically dispersed Fe–Pt dual-site catalysts for efficient oxygen reduction reactions, Chemical Science, 14 (2023) 3277-3285.
[18] S. Liu, H. Xu, D. Liu, H. Yu, F. Zhang, P. Zhang, R. Zhang, W. Liu, Identify the Activity Origin of Pt Single-Atom Catalyst via Atom-by-Atom Counting, Journal of the American Chemical Society, 143 (2021) 15243-15249.
[19] T. Lin, Y. Shen, M. Ge, Y. Li, Z. Jiang, Z.H. Lyu, J. Liu, L. Gu, X. Liu, Atomic-scale Observation of the Generation and Dispersion of Iron Single Atoms, Chemical Research in Chinese Universities, (2025).
[20] R.F. Egerton, Electron energy-loss spectroscopy in the electron microscope, Springer Science & Business Media, 2011.
[21] E. Rutherford, The scattering of α and β particles by matter and the structure of the atom, Philosophical Magazine Series 6, 21 (1911) 669-688.
[22] S.J. Pennycook, D.E. Jesson, High-resolution Z-contrast imaging of crystals, Ultramicroscopy, 37 (1991) 14-38.
[23] J.M. Cowley, A.F. Moodie, The scattering of electrons by atoms and crystals. I. A new theoretical approach, Acta Crystallographica, 10 (1957) 609-619.
[24] G. Behan, E.C. Cosgriff, A.I. Kirkland, P.D. Nellist, Three-dimensional imaging by optical sectioning in the aberration-corrected scanning transmission electron microscope, Phil. Trans. R. Soc. A., 367 (2009) 3825-3844.
[25] F. Hosokawa, T. Shinkawa, Y. Arai, T. Sannomiya, Benchmark test of accelerated multi-slice simulation by GPGPU, Ultramicroscopy, 158 (2015) 56-64.
[26] E.J. Kirkland, Advanced computing in electron microscopy, Springer Nature, 2020.
[27] J. Madsen, T. Susi, The abTEM code: transmission electron microscopy from first principles, Open Research Europe, 1 (2021) 24.
[28] C.A. Liu, Y. Cui, Y. Zhou, The recent progress of single-atom catalysts on amorphous substrates for electrocatalysis, Energy Materials, 5 (2025) 500001.
[29] J.B. Xi, H.Y. Sun, D. Wang, Z.Y. Zhang, X.M. Duan, J.W. Xiao, F. Xiao, L.M. Liu, S. Wang, Confined-interface-directed synthesis of Palladium single-atom catalysts on graphene/amorphous carbon, Applied Catalysis B: Environmental, 225 (2018) 291-297.
[30] P.K. Chu, L.H. Li, Characterization of amorphous and nanocrystalline carbon films, Materials Chemistry and Physics, 96 (2006) 253-277.
[31] N. Laanait, M. Ziatdinov, Q. He, A. Borisevich, Identifying local structural states in atomic imaging by computer vision, Advanced Structural and Chemical Imaging, 2 (2016) 14.
[32] G. Wang, C. Lopez-Molina, B. De Baets, Automated blob detection using iterative Laplacian of Gaussian filtering and unilateral second-order Gaussian kernels, Digital Signal Processing, 96 (2020)

102592.

[33] D. Sun, S. Roth, M.J. Black, Secrets of optical flow estimation and their principles, in: 2010 IEEE Computer Society Conference on Computer Vision and Pattern Recognition, 2010, pp. 2432-2439.

[34] X. Zeng, J. Shui, X. Liu, Q. Liu, Y. Li, J. Shang, L. Zheng, R. Yu, Single-Atom to Single-Atom Grafting of Pt1 onto FeN4 Center: Pt1@FeNC Multifunctional Electrocatalyst with Significantly Enhanced Properties, Advanced Energy Materials, 8 (2018) 1701345.

[35] F. Xiao, Q. Wang, G.L. Xu, X. Qin, I. Hwang, C.J. Sun, M. Liu, W. Hua, H.w. Wu, S. Zhu, J.C. Li, J.G. Wang, Y. Zhu, D. Wu, Z. Wei, M. Gu, K. Amine, M. Shao, Atomically dispersed Pt and Fe sites and Pt–Fe nanoparticles for durable proton exchange membrane fuel cells, Nature Catalysis, 5 (2022) 503-512.

[36] K.A. Mkhoyan, S.E. Maccagnano-Zacher, E.J. Kirkland, J. Silcox, Effects of amorphous layers on ADF-STEM imaging, Ultramicroscopy, 108 (2008) 791-803.

[37] A. Mittal, K.A. Mkhoyan, Limits in detecting an individual dopant atom embedded in a crystal, Ultramicroscopy, 111 (2011) 1101-1110.

[38] M. Takeguchi, A. Hashimoto, K. Mitsuishi, Depth sectioning using environmental and atomic-resolution STEM, Microscopy, 73 (2024) 145-153.

[39] R. Ishikawa, A.R. Lupini, Y. Hinuma, S.J. Pennycook, Large-angle illumination STEM: Toward three-dimensional atom-by-atom imaging, Ultramicroscopy, 151 (2015) 122-129.

# Supporting Information

## Differentiation of Distinct Single Atoms via Multi-Defocus Fusion Method

Yangfan Li [a,b], Yue Pan [a,b], Xincheng Lei [a,b], Weiwei Chen [a,b], Yang Shen [a,b], Mengshu Ge [a], Xiaozhi Liu [a], Dong Su[a,b]

[a]Beijing National Laboratory for Condensed Matter Physics, Institute of Physics, Chinese Academy of Sciences, Beijing, 100190, China

[b] School of Physical Sciences, University of Chinese Academy of Sciences, Beijing, 100049, China

### 1. Simulation parameters

Simulations were performed using the abTEM code[1], employing the multi-slice approach described in the main text (Section 2.1). We simulated images under different defocus conditions as well as with varying convergent and collection angles. **Table S1** lists the electron microscopy parameters used in the simulation work of this study.

**Table S1**. Parameters used in the simulation.

| Parameter | Value |
| --- | --- |
| Acceleration voltage | 200 kV |
| Convergence angle | 20 mrad, 28 mrad, 36 mrad |
| Collection angle | 45-250 mrad, 54-250 mrad, 68-250 mrad |
| Spherical aberration | 500 nm |

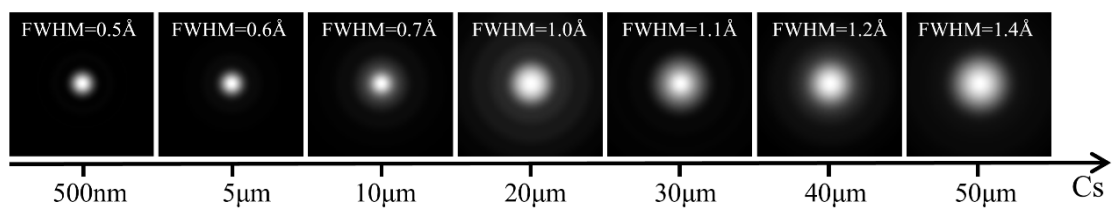

**Fig. S1.** Simulated HAADF-STEM images of single Fe atoms in vacuum under different spherical aberration coefficients (Cs). Each image has a side length of 4 Å and is individually normalized to demonstrate the effect.

## 2. Thickness Measurement of amorphous carbon support

In order to measure the carbon support thickness, electron Energy Loss Spectroscopy (EELS) is employed in HAADF-STEM experiments. The thickness measurement with EELS is fundamentally based on evaluating the inelastic mean free path (MFP) of electrons traversing through the specimen. When electrons pass through a thin specimen, a fraction of the electrons undergoes inelastic scattering events, resulting in characteristic energy loss signals. According to the Beer–Lambert law for electron scattering[2], the thickness of the specimen ($t$) can be quantitatively correlated with the intensity ratio of the inelastically scattered electrons ($I_{inelastic}$) to the total transmitted electrons ($I_{total}$) as follows[3]:

$$t = \lambda ln\left(\frac{I_{total}}{I_{elastic}}\right) = \lambda ln\left(\frac{I_{elastic} + I_{inelastic}}{I_{elastic}}\right) \quad (1)$$

where $\lambda$ represents the electron inelastic mean free path, $I_{elastic}$ denotes the intensity of elastically scattered electrons (zero-loss peak intensity), and $I_{inelastic}$ denotes the intensity of inelastically scattered electrons (energy-loss electrons).

In our study, EELS measurements were conducted using a selected flat and uniform region of the amorphous carbon support in the STEM mode. EELS spectra were acquired by positioning the electron beam precisely on the region of interest, ensuring stable and reproducible measurements. The resulting EELS data contain both zero-loss peaks and characteristic inelastic energy-loss signals, from which we derived thickness maps following the aforementioned theoretical relation.

**Fig. S2** summarizes the results from our EELS-based thickness measurement procedure. Specifically, **Fig. S2(a)** presents a representative EELS spectrum acquired from the selected amorphous carbon support area. From this spectrum, the zero-loss peak and the inelastic scattering region are clearly distinguishable, allowing for accurate extraction of intensity information needed for thickness calculation. To better visualize and quantify the thickness distribution, **Fig. S2(b)** presents a histogram derived from the thickness map. The histogram indicates a relatively narrow thickness distribution, confirming the uniformity of the selected amorphous carbon support. From this statistical analysis, we determine that the average thickness of the amorphous carbon support used in this study is approximately 7 nm. Therefore, we employed a 7 nm thick amorphous carbon support model in all subsequent STEM image simulations.

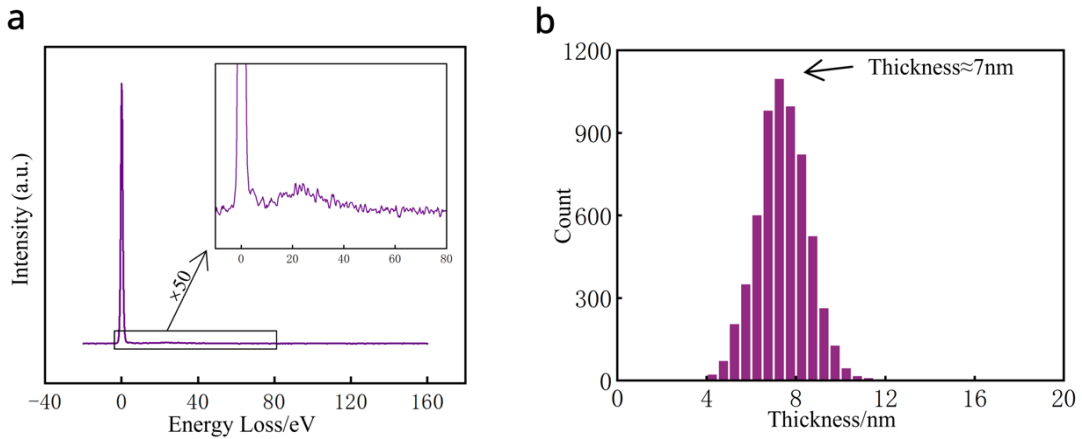

**Fig. S2**. Electron energy-loss spectroscopy (EELS)-based procedure for the thickness measurement of amorphous carbon support: (a) Representative EELS spectrum showing the distinct zero-loss peak and inelastic scattering region used for thickness calculation; (b) Histogram of support thickness derived from the thickness map, confirming a narrow and uniform thickness distribution.

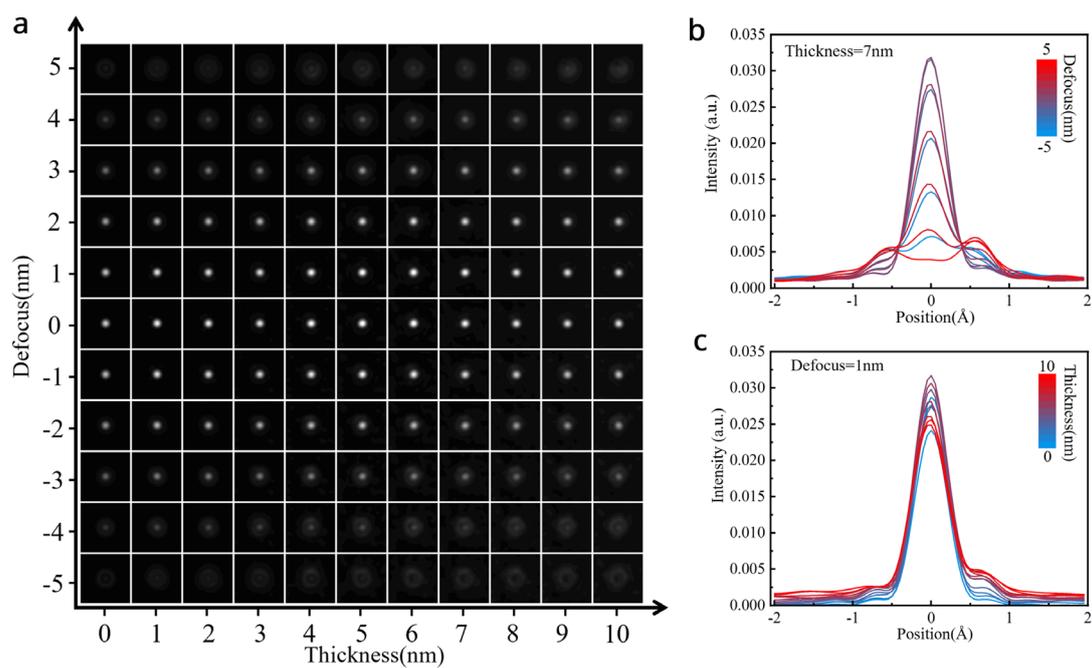

**Fig. S3**. HAADF-STEM image simulation of single Pt atom on amorphous carbon support. (a) Simulated map images obtained under various amorphous carbon thicknesses (0–10 nm) and defocus (−5 to +5 nm) conditions. (b) Line profiles recorded at a fixed carbon thickness of 7 nm under different defocus values. (c) Line profiles recorded at a fixed defocus of 1 nm for amorphous carbon support of varying thickness.

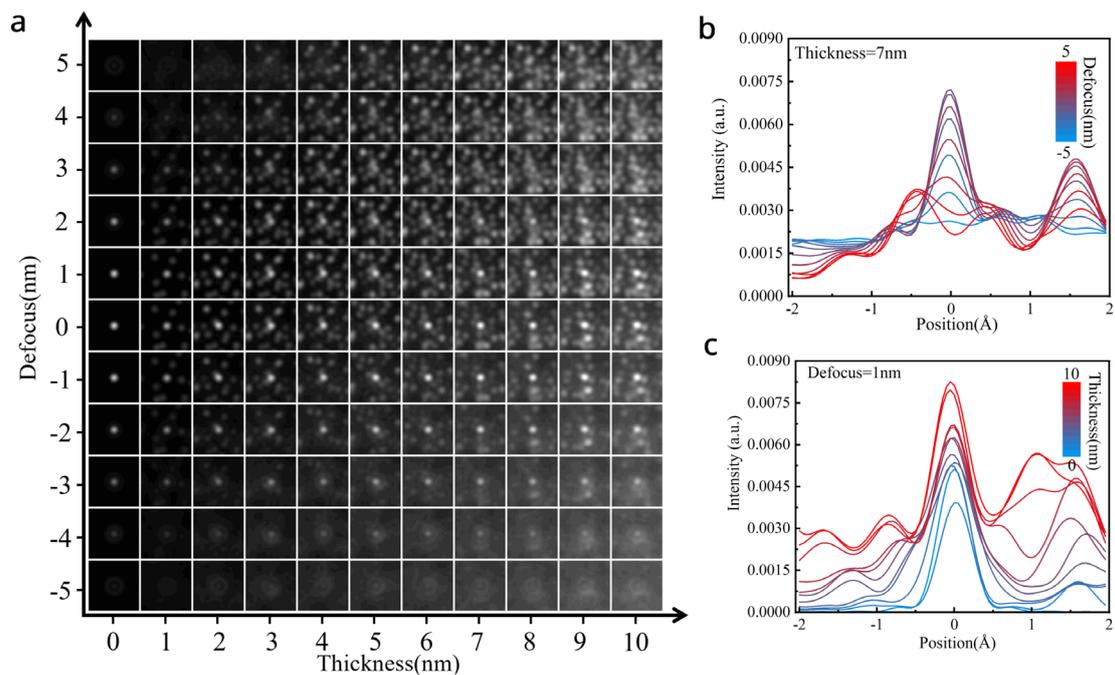

**Fig. S4**. HAADF-STEM image simulation of single Fe atoms on amorphous silicon support. (a) Simulated map images obtained under various silicon thicknesses (0–10 nm) and defocus (−5 to +5 nm) conditions. (b) Line profiles recorded at a fixed silicon thickness of 7 nm under different defocus values. (c) Line profiles recorded at a fixed defocus of 1 nm for amorphous silicon support of varying thickness.

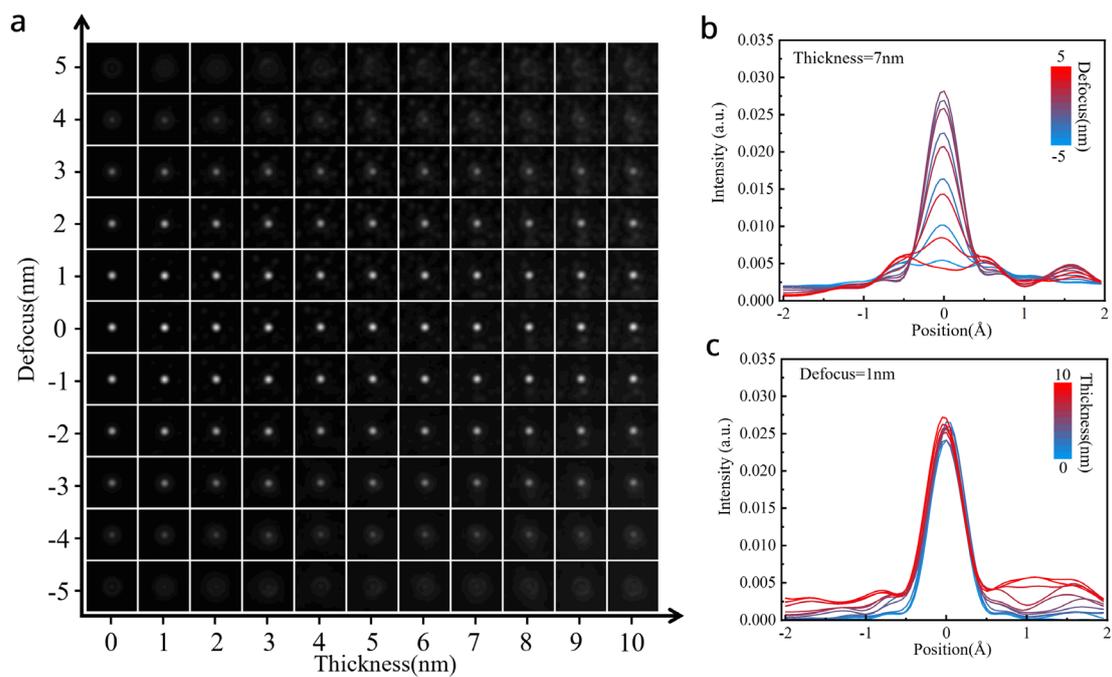

**Fig. S5**. HAADF-STEM image simulation of single Pt atom on amorphous silicon support. (a) Simulated map images obtained under various amorphous silicon thicknesses (0–10 nm) and defocus (−5 to +5 nm) conditions. (b) Line profiles recorded at a fixed silicon thickness of 7 nm under different defocus values. (c) Line profiles recorded at a fixed defocus of 1 nm for amorphous silicon supports of varying thickness.

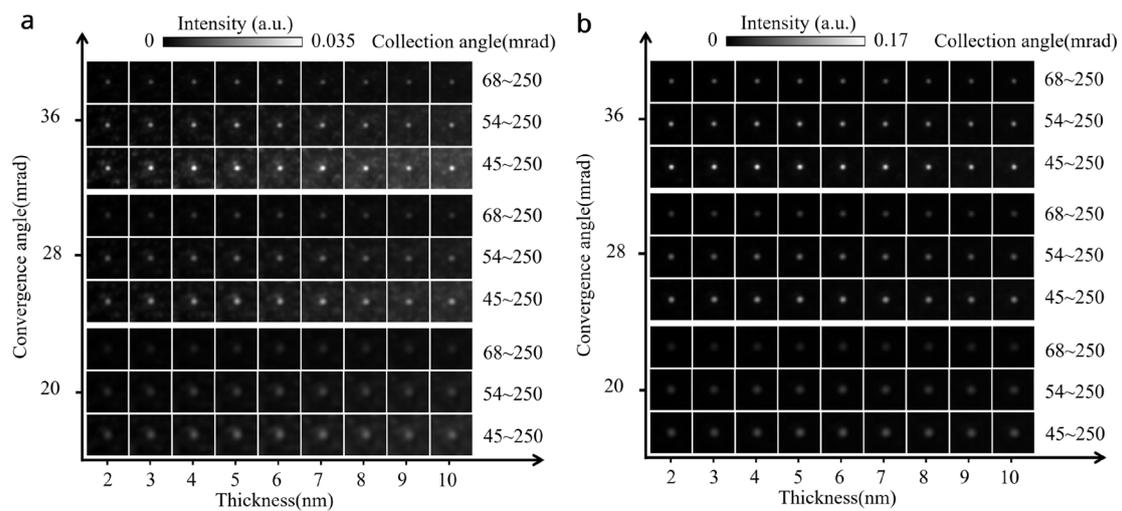

**Fig. S6**. HAADF-STEM image simulation at various convergence/collection angles. (a) Original simulated images of a single Fe atom on amorphous carbon with different convergence and collection angles. (b) Original simulated images of a single Pt atom on amorphous carbon under the same series of imaging conditions.

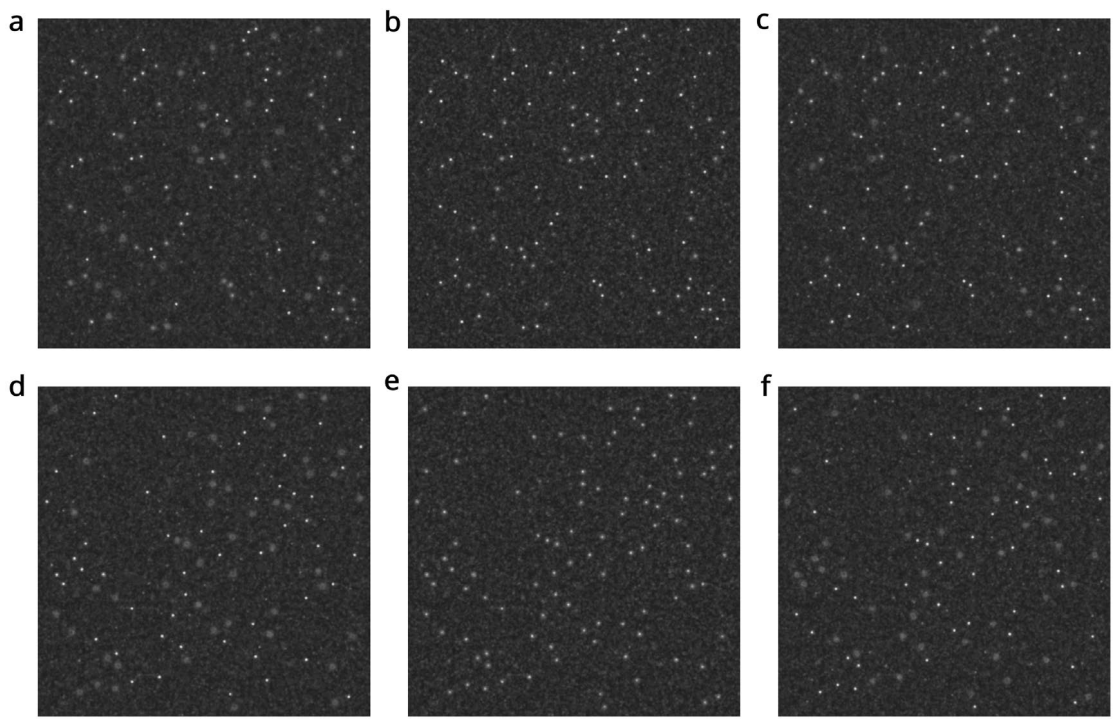

**Fig. S7**. Original (non-inverted contrast) simulated HAADF-STEM images of single Fe and Pt atoms supported within/on a 7 nm-thick amorphous carbon support under different defocus conditions. (a–c) Uniform internal atomic distribution model, corresponding directly to Fig. 4(b–d), showing images simulated at upper surface focus (a), middle plane focus (b), and lower surface focus (c). (d–f) Surface atomic distribution model, corresponding directly to Fig. 4(j–l), showing images simulated at upper surface focus (d), middle plane focus (e), and lower surface focus (f). Simulation parameters: convergence semi-angle: 28 mrad; detector collection semi-angle: 54–250 mrad.

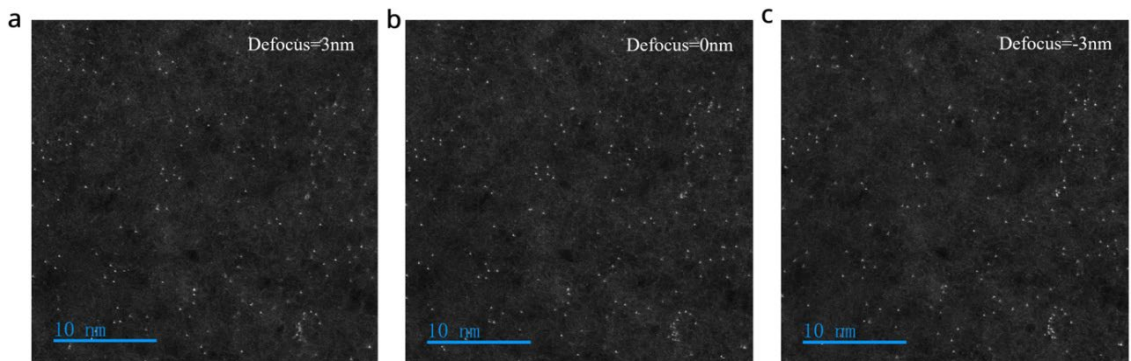

**Fig. S8**. Experimental HAADF-STEM images of Fe/Pt single atoms dispersed on ultrathin amorphous carbon support, acquired under nominal defocus of (a) +3 nm, (b) 0 nm, and (c) −3 nm.

**References**

[1] J. Madsen, T. Susi, The abTEM code: transmission electron microscopy from first principles, Open Research Europe, 1 (2021) 24.

[2] T. Malis, S.C. Cheng, R.F. Egerton, EELS log-ratio technique for specimen-thickness measurement in the TEM, Journal of electron microscopy technique, 8 (1988) 193-200.

[3] R.F. Egerton, Electron energy-loss spectroscopy in the electron microscope, Springer Science & Business Media, 2011.